\documentstyle[preprint,aps,epsfig,here]{revtex}
\begin{document}
 
\title{\bf 
With four Standard Model families, the LHC could discover the Higgs 
boson with
a few fb$^{-1}$
}

\author{E. Ar{\i}k$^{a}$, M. Ar{\i}k$^{a}$, S.A. {\c C}etin$^{a}$, T. {\c
C}onka$^{a}$, A.
Mailov$^{a,b}$, S. Sultansoy$^{b,c}$}

\address{ $^a$Department~of~ Physics,~ Faculty~ of~ Arts~ and~ Sciences,
~Bo\u{g}azi\c{c}i~ University,~ 80815,~ Bebek,\\ {\. I}stanbul,~ Turkey\\
 $^b$Institute~ of~ Physics,~ Academy~ of~ Sciences,~ H.~ Cavid~ Avenue,~
370143,\\
Baku,~ Azerbaijan\\
 $^c$Department~ of~ Physics,~ Faculty~ of~ Arts~ and~ Sciences,~ Gazi~
University,~ 06500,~Teknikokullar,\\
 Ankara,~Turkey} 
\maketitle

\vspace*{1.0cm}

\begin{abstract}

The existence of a 4$^{th}$ SM family 
would produce a large
enhancement of
the gluon fusion channel of Higgs boson production at hadron colliders. In
this case, the SM Higgs boson could be seen at the CERN Large Hadron
Collider (LHC) via the golden mode ($H^0 \rightarrow 4l$) 
with an integral luminosity of only a few fb$^{-1}$.

\end{abstract}
\vskip 1.0cm

\normalsize 
\baselineskip=20pt

One of the primary goals at the LHC is to observe the Standard Model (SM)
Higgs particle ($H^0$).
The LEP data already set a lower bound on the Higgs boson mass at
114.1 GeV \cite{lep}. Proton-proton collisions at a center of mass 
energy of 14 TeV at the LHC 
will produce $H^0$, mainly through gluon-gluon
fusion \cite{tp}, with a mass up to 1 TeV. 

The main contribution to the production of the Higgs boson in the three SM
family case, comes from the well
known triangular diagram with a t-quark loop. If the $4^{th}$ SM family
\cite{4th1,4th2} exists, we have two additional diagrams where the t-quark is
replaced by $u_4$ and $d_4$ quarks. 

In order to see the effect of $u_{4}$ and $d_{4}$ on the $gg
\rightarrow H^0$ process, the decay width of $H^0 \rightarrow gg$ is
calculated using the following formula \cite{collider}:

\begin{eqnarray}
\Gamma (H \rightarrow gg) = \frac{G_{F} m_{H}^{3}}{36 \sqrt{2} \pi} 
\left(\frac{\alpha_{s}(m_{H}^{2})}{\pi} \right)^{2} |I|^{2}~~. 
\end{eqnarray}
The quantity $I$ is expressed as $I = \Sigma I_{q} = I_t + I_{u_4} + I_{d_4}$, where
$I_{q}$ is given by $I_{q} = 3[2\lambda_{q} + \lambda_{q}(4\lambda_{q}
-1)f(\lambda_{q})]$, and
$\lambda_{q} = (m_{q} / m_{H})^{2}$.
Here the function $f(\lambda_{q})$ takes two different forms:
\begin{eqnarray}
\hspace{3.8cm} f(\lambda_{q}) = -2 \left( arcsin \frac{1}{2 \sqrt{\lambda_{q}}}
\right)^{2},
\hspace{1.cm} for\hspace{0.2cm}  \lambda_{q}>\frac{1}{4}~~,
\end{eqnarray}
\begin{eqnarray}
\hspace{2.6cm} f(\lambda_{q}) = \frac{1}{2} \left(ln\frac{\eta^{+}}{\eta^{-}}
\right)^{2} - \frac{\pi^{2}}{2} - i\pi ln\frac{\eta^{+}}{\eta^{-}}~~,
\hspace{1.14cm} for\hspace{0.2cm}  \lambda_{q}<\frac{1}{4}~~,
\end{eqnarray}
where $\eta^{\pm} = \frac{1}{2} \pm \sqrt{\frac{1}{4} - \lambda_{q}}$~~. 
\\

Figure 1 shows the enhancement factor $K$ of the $H^0$ production
cross-section via gluon-gluon fusion defined as
\begin{eqnarray}
K~=~\frac{|I_{t}+I_{u_{4}}+I_{d_{4}}|^{2}}{|I_{t}|^{2}}~.
\end{eqnarray}
In this figure, the $4^{th}$ family quark masses $m_{u_4}$=$m_{d_4}$=320
and 640 GeV are considered \cite{4th2}. In addition,
$m_{u_4}$=$m_{d_4}$=200 GeV case is shown as an illustration.

This enhancement will improve the signal of $H^0$ in all decay modes
\cite{note}. In our opinion, the most promising channel is the so-called
"golden-mode" where Higgs boson decays into four charged leptons. 
For example \cite{orhan},
if the 4$^{th}$ SM
family exists,
the upgraded Tevatron, with an integral luminosity of 15 fb$^{-1}$ 
per experiment and
with $\sim 25\%$ overall detector acceptance, 
may already get indications of the Higgs boson via 
the golden mode with significance of 2-3 $\sigma$ 
for 175$<$m$_H$$<$300 GeV. 

Since the existence of the $4^{th}$ SM family provides an increase in 
$H^0$ production, and the background will not be affected, 
the significance of the signal at the LHC will be improved. 
In the 3 family case, the most promising decay channels for the detection
of the SM Higgs boson at the LHC-ATLAS experiment are \cite{tp}: 
$H^0 \rightarrow \gamma \gamma$ ($m_H$ = 80 - 120 GeV), $H^0 \rightarrow
ZZ^{*},ZZ \rightarrow 4l$ ($m_H$ =
130 - 800 GeV) and $H^0 \rightarrow ZZ,W^{\pm}W^{\mp} \rightarrow
lljj,ll\nu\nu,l\nu jj $ ($m_H$ = 500 - 1000 GeV); here $l$ denotes
$e$ or $\mu$, and $j$ denotes jet.

In this work, only the $H^0 \rightarrow 4l$ (golden mode) channel is
considered. The signal (S) and background (B) events
expected in ATLAS \cite{tdr2} in the 3-family case with 30 $fb^{-1}$ 
are given in Table 1. The statistical significance (SS) values are 
calculated using Poisson statistics and expressed in terms of 
Gaussian $\sigma$-units.
Both here and in \cite{tdr2} the analysis cuts are less restrictive for 
$m_H \ge 200 GeV$, where $H^0$ decays to two real $Z$.

The $H^0 \rightarrow ZZ^{(*)},WW$ decay processes do not have any quark
loops, hence the decay widths are not affected by the $4^{th}$ family.
The enhancement factor $K$ will increase the total decay width of $H^0$ 
and slightly decrease the Branching Ratios (BR) of $WW$ and $ZZ$
modes \cite{note}. In Table 2, the signal, background and significance
values of $H^0 \rightarrow 4l$ are given in the presence of the $4^{th}$
SM family, by extrapolating the values given in Table 1 for different
luminosities. In the calculations, $m_{u_4}$=$m_{d_4}$=320 GeV is assumed. 
Figure 1 shows that using $m_{u_4}$=$m_{d_4}$=640 GeV does not change the
results much up to $m_H$=600 GeV .

As seen in Table 2, the existence of the $4^{th}$ family will result 
in a very conspicuous improvement of the significance of the golden mode signal
at $L_{int}$=30 fb$^{-1}$. Moreover, almost all of the Higgs
boson mass region will be covered even at $L_{int}$ =
3 fb$^{-1}$. This situation is clearly demonstrated in
Figure 2, which represents the integral luminosities needed to achieve a
$5\sigma$
significance level.

Finally, in the presence of the $4^{th}$ SM family, the upgraded Tevatron
would get indications of the Higgs boson via the golden mode for 
175$<$m$_H$$<$300 GeV when an
integrated luminosity of 15 $fb^{-1}$ per experiment is reached. On the other hand, at the
LHC, the region of $125 \le m_H \le 600$ GeV would be fully covered with
5$\sigma$ or more significance even at $L_{int}$ =
3 fb$^{-1}$.

We are grateful to our colleagues in ATLAS Collaboration and especially to O. {\c
C}ak{\i}r for useful discussions.

\pagebreak

\begin{center}
TABLES
\end{center}

\small{TABLE I. Signal (S), background (B) and statistical 
significance (SS)
values for the $pp \rightarrow H^0+X,~ H^0 \rightarrow 4l$ channel with
30 fb$^{-1}$ in 3 SM family case.}
\label{t:}
\vskip 0.5 cm
\begin{center}
\begin{tabular}{c|c|c|c|c|c|c|c|c|c|c|c|c|c}
\hline   
$m_H(GeV)$&~120~&~130~&~150~&~170~&~180~&~200~&~240~&~280~&~320~&~360~&~400~&~500~&~600~
\\ \hline
$S$&4.1&11.4&26.8&7.6&19.7&134.0&127.0&110.0&105.0&105.0&86.0&44.0&23.0
\\ \hline
$B$&1.4 &2.6&3.0&3.1&3.1&74.0&57.0&43.0&33.0&29.0&29.0&17.0&15.0
\\ \hline
$SS$&2.4&4.8&$>$~8&3.2&7.1&$>$~8&$>$~8&$>$~8&$>$~8&$>$~8&$>$~8&$>$~8&4.9
\\ \hline  
\end{tabular}
\end{center} 
\vskip 1.5 cm

\small{TABLE II. Signal (S), background (B) and statistical 
significance (SS)
values for the $pp \rightarrow H^0+X,~ H^0 \rightarrow 4l$ channel in 
the 4 SM family case ($m_{4}$=320 GeV) for various luminosities.}
\vskip 0.5 cm
{\centering \begin{tabular}{c||c|c|c||c|c|c}
\hline   
\multicolumn{1}{c||}{$m_H$}&\multicolumn{3}{c||}{$L_{int}$=30 
$fb^{-1}$}&\multicolumn{3}{c}{$L_{int}$=3 
$fb^{-1}$}
\\ \cline{2-7}
$(GeV)$&$~~S~~$&$~~B~~$&$~SS~$&$~~S~~$&$~~B~~$&$~SS~$
\\ \hline
120 &25.1&1.4&$>$~8&2.5&0.1&3.2
\\ \hline
130 &71.4&2.6&$>$~8&7.1&0.3&5.6
\\ \hline  
150  &197.0&3.0&$>$~8&19.7&0.3&$>$~8
\\ \hline  
170  &64.0&3.1&$>$~8&6.4&0.3&5.2
\\ \hline  
180  &165.1&3.1&$>$~8&16.5&0.3&$>$~8
\\ \hline  
200  &1106.8&74.0&$>$~8&110.7&7.4&$>$~8
\\ \hline  
240  &1005.8&57.0&$>$~8&100.6&5.7&$>$~8
\\ \hline  
280  &819.5&43.0&$>$~8&82.0&4.3&$>$~8
\\ \hline  
320  &711.9&33.0&$>$~8&71.2&3.3&$>$~8
\\ \hline  
360  &578.6&29.0&$>$~8&57.8&2.9&$>$~8
\\ \hline  
400  &422.2&29.0&$>$~8&42.2&2.9&$>$~8
\\ \hline  
500  &225.3&17.0&$>$~8&22.5&1.7&$>$~8
\\ \hline  
600  &148.1&15.0&$>$~8&14.8&1.5&6.8
\\ \hline  
\end{tabular}\par}

\begin{figure}[H]
\epsfig{figure=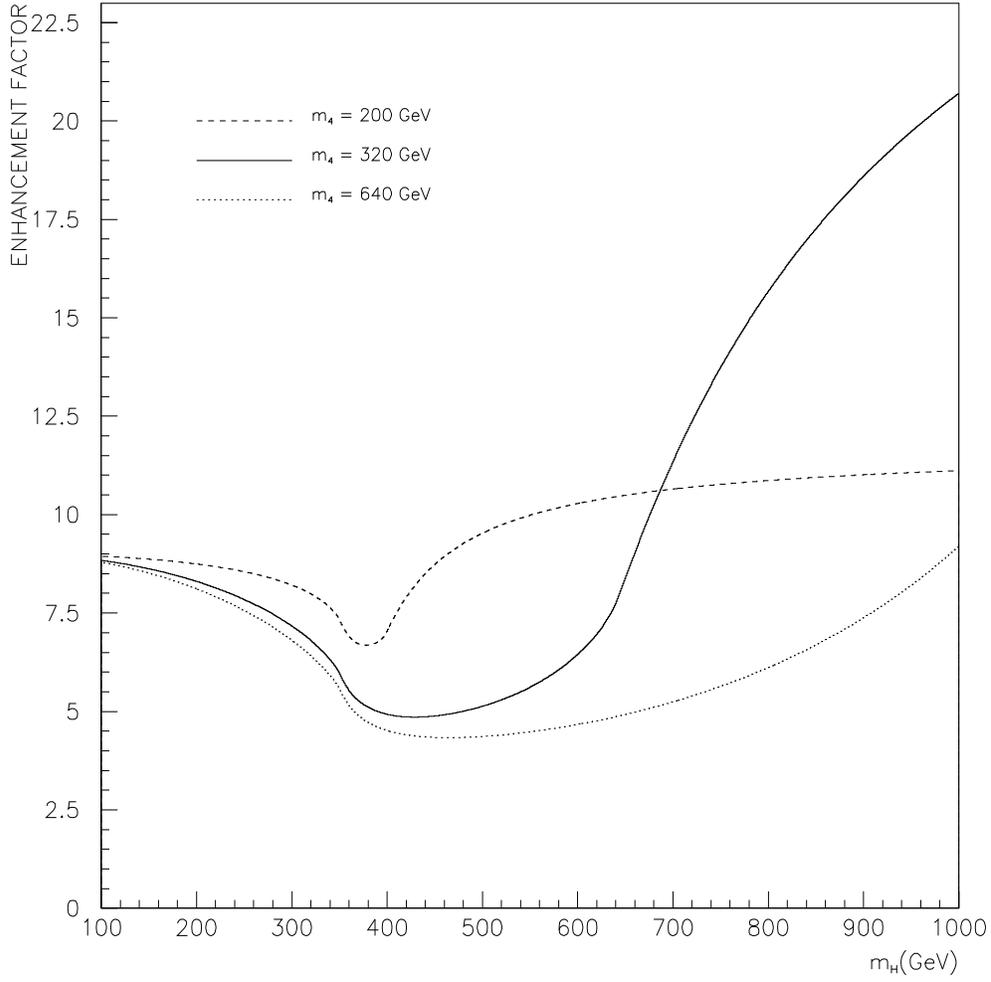,height=13.0cm,width=13.0cm}
\vspace*{0.5cm}
\caption{Enhancement factor of the Higgs boson production ($gg
\rightarrow H^0$) due to the $4^{th}$ SM family quarks.}
\end{figure}

\begin{figure}[H]
\epsfig{figure=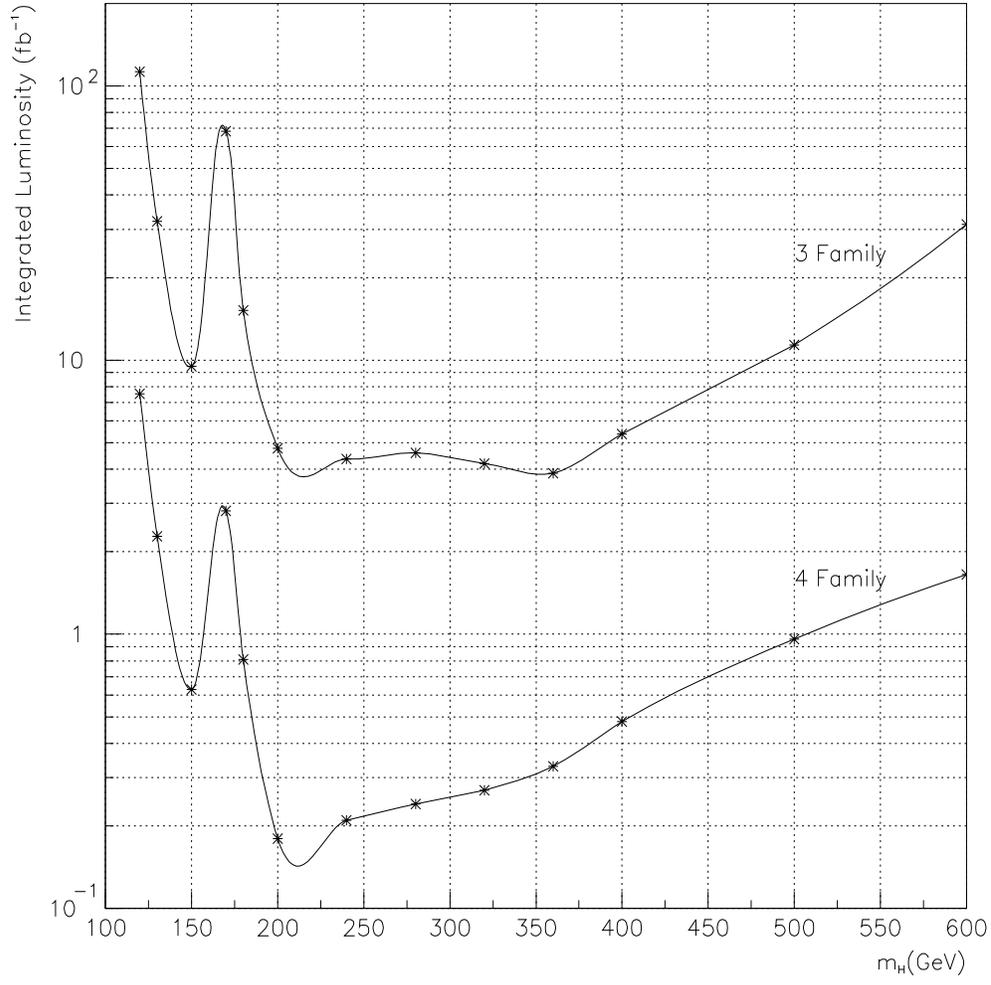,height=13.0cm,width=13.0cm}
\vspace*{0.5cm}
\caption{LHC luminosity values corresponding to 5$\sigma$ significance
level of the golden mode signal for 3 and 4 family cases.}
\end{figure}

\end{document}